\documentclass{article}

\PassOptionsToPackage{numbers, compress}{natbib}
\usepackage[final]{nips_2016}

\usepackage[utf8]{inputenc} 
\usepackage[T1]{fontenc}    
\usepackage{hyperref}       
\usepackage{url}            
\usepackage{booktabs}       
\usepackage{amsfonts}       
\usepackage{nicefrac}       
\usepackage{microtype}      
\usepackage{amssymb,array,latexsym,amsmath,graphics,graphicx,array}
\usepackage{algorithm, algorithmicx, algpseudocode}
\title{Automated scalable segmentation of neurons from multispectral images}

\author{
  Uygar S\"umb\"ul \\
  Grossman Center for the Statistics of Mind\\
  and Dept. of Statistics, Columbia University\\
  \And
  Douglas Roossien Jr. \\
  University of Michigan Medical School \\
  \And
  Fei Chen \\
  MIT Media Lab and McGovern Institute \\
  \And
  Nicholas Barry \\
  MIT Media Lab and McGovern Institute \\
  \And
  Edward S. Boyden \\
  MIT Media Lab and McGovern Institute \\
  \And
  Dawen Cai \\
  University of Michigan Medical School \\
  \And
  John P. Cunningham \\
  Grossman Center for the Statistics of Mind \\
  and Dept. of Statistics, Columbia University \\
  \And
  Liam Paninski \\
  Grossman Center for the Statistics of Mind \\
  and Dept. of Statistics, Columbia University \\
}

\begin{document}

\maketitle

\begin{abstract}
Reconstruction of neuroanatomy is a fundamental problem in neuroscience. Stochastic expression of colors in individual cells is a promising tool, although its use in the nervous system has been limited due to various sources of variability in expression. Moreover, the intermingled anatomy of neuronal trees is challenging for existing segmentation algorithms. Here, we propose a method to automate the segmentation of neurons in such (potentially pseudo-colored) images. The method uses spatio-color relations between the voxels, generates supervoxels to reduce the problem size by four orders of magnitude before the final segmentation, and is parallelizable over the supervoxels. To quantify performance and gain insight, we generate simulated images, where the noise level and characteristics, the density of expression, and the number of fluorophore types are variable. We also present segmentations of real Brainbow images of the mouse hippocampus, which reveal many of the dendritic segments.
\end{abstract}

\vspace{-.5cm}
\section{Introduction}
\label{intro}
\vspace{-.3cm}
Studying the anatomy of individual neurons and the circuits they form is a classical approach to understanding how nervous systems function since Ram\'on y Cajal's founding work. Despite a century of research, the problem remains open due to a lack of technological tools: mapping neuronal structures requires a large field of view, a high resolution, a robust labeling technique, and computational methods to sort the data. Stochastic labeling methods have been developed to endow individual neurons with color tags~\cite{livet2007transgenic, marblestone2014rosetta}. This approach to neural circuit mapping can utilize the light microscope, provides a high-throughput and the potential to monitor the circuits over time, and complements the dense, small scale connectomic studies using electron microscopy~\cite{takemura2013visual} with its large field-of-view. However, its use has been limited due to its reliance on manual segmentation.

The initial stochastic, spectral labeling (Brainbow) method had a number of limitations for neuroscience applications including incomplete filling of neuronal arbors, disproportionate expression of the nonrecombined fluorescent proteins in the transgene, suboptimal fluorescence intensity, and color shift during imaging. Many of these limitations have since improved~\cite{cai2013improved} and developments in various aspects of light microscopy provide further opportunities~\cite{chen2015expansion,chung2013clarity,betzig2006imaging,rust2006sub}. Moreover, recent approaches promise a dramatic increase in the number of (pseudo) color sources~\cite{zador2012sequencing,lee2014highly,chen2015spatially}. Taken together, these advances have made light microscopy a much more powerful tool for neuroanatomy and connectomics. However, existing automated segmentation methods are inadequate due to the spatio-color nature of the problem, the size of the images, and the complicated anatomy of neuronal arbors. Scalable methods that take into account the high-dimensional nature of the problem are needed.

Here, we propose a series of operations to segment 3-D images of stochastically tagged nervous tissues. Fundamentally, the computational problem arises due to insufficient color consistency within individual cells, and the voxels occupied by more than one neuron. We denoise the image stack through collaborative filtering~\cite{maggioni2013nonlocal}, and obtain a supervoxel representation that reduces the problem size by four orders of magnitude. We consider the segmentation of neurons as a graph segmentation problem~\cite{von2007tutorial}, where the nodes are the supervoxels. Spatial discontinuities and color inhomogeneities within segmented neurons are penalized using this graph representation. While we concentrate on neuron segmentation in this paper, our method should be equally applicable to the segmentation of other cell classes such as glia.

To study various aspects of stochastic multispectral labeling, we present a basic simulation algorithm that starts from actual single neuron reconstructions. We apply our method on such simulated images of retinal ganglion cells, and on two different real Brainbow images of hippocampal neurons, where one dataset is obtained by expansion microscopy~\cite{chen2015expansion}.

\vspace{-.3cm}
\section{Methods}
\begin{figure}
\centering
\includegraphics*[viewport = 0 0 2450 2450, width=0.8\textwidth]{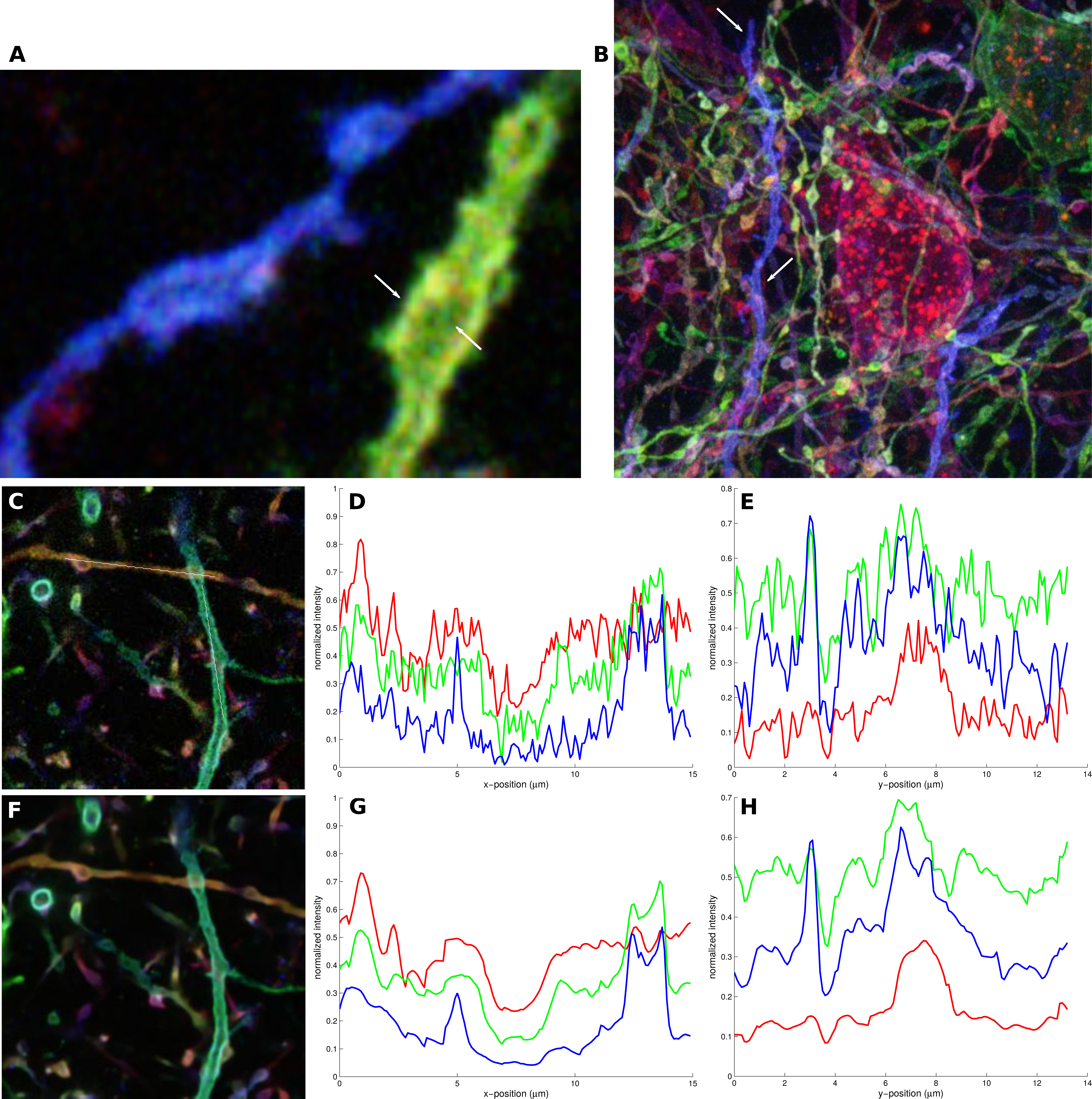}
\caption{{\bf Multiple noise sources affect the color consistency in Brainbow images.} {\bf a,} An $85\times 121$ Brainbow image patch from a single slice (physical size: $8.5\mu\times 12.1\mu$). Expression level differs significantly between the membrane and the cytoplasm along a neurite (arrows). {\bf b,} A maximum intensity projection view of the $3$-d image stack. Color shifts along a single neurite, which travels to the top edge and into the page (arrows). {\bf c,} A $300\times 300$ image patch from a single slice of a different Brainbow image (physical size: $30\mu\times 30\mu$). {\bf d,} The intensity variations of the different color channels along the horizontal line in {\bf c}. {\bf e,} Same as {\bf d} for the vertical line in {\bf c}. {\bf f,} The image patch in {\bf c} after denoising. {\bf g--h,} Same as {\bf d} and {\bf e} after denoising. For the plots, the range of individual color channels is $[0,1]$.}
\label{noiseSources}
\end{figure}

\vspace{-.3cm}
Successful segmentations of color-coded neural images should consider both the connected nature of neuronal anatomy and the color consistency of the Brainbow construct. However, the size and the noise level of the problem prohibit a voxel-level approach (Fig.~\ref{noiseSources}). Methods that are popular in hyperspectral imaging applications, such as nonnegative matrix factorization~\cite{lee2001algorithms}, are not immediately suitable either because the number of color channels are too few and it is not easy to model neuronal anatomy within these frameworks. Therefore, we develop (i) a supervoxelization strategy, (ii) explicitly define graph representations on the set of supervoxels, and (iii) design the edge weights to capture the spatio-color relations (Fig.~\ref{pipeline2}a).

\vspace{-.3cm}
\subsection{Denoising the image stack}
\label{denoising}
\begin{figure}
  \begin{minipage}[l]{0.8\textwidth}
    \includegraphics*[viewport = 0 0 5200 5600, width=\textwidth]{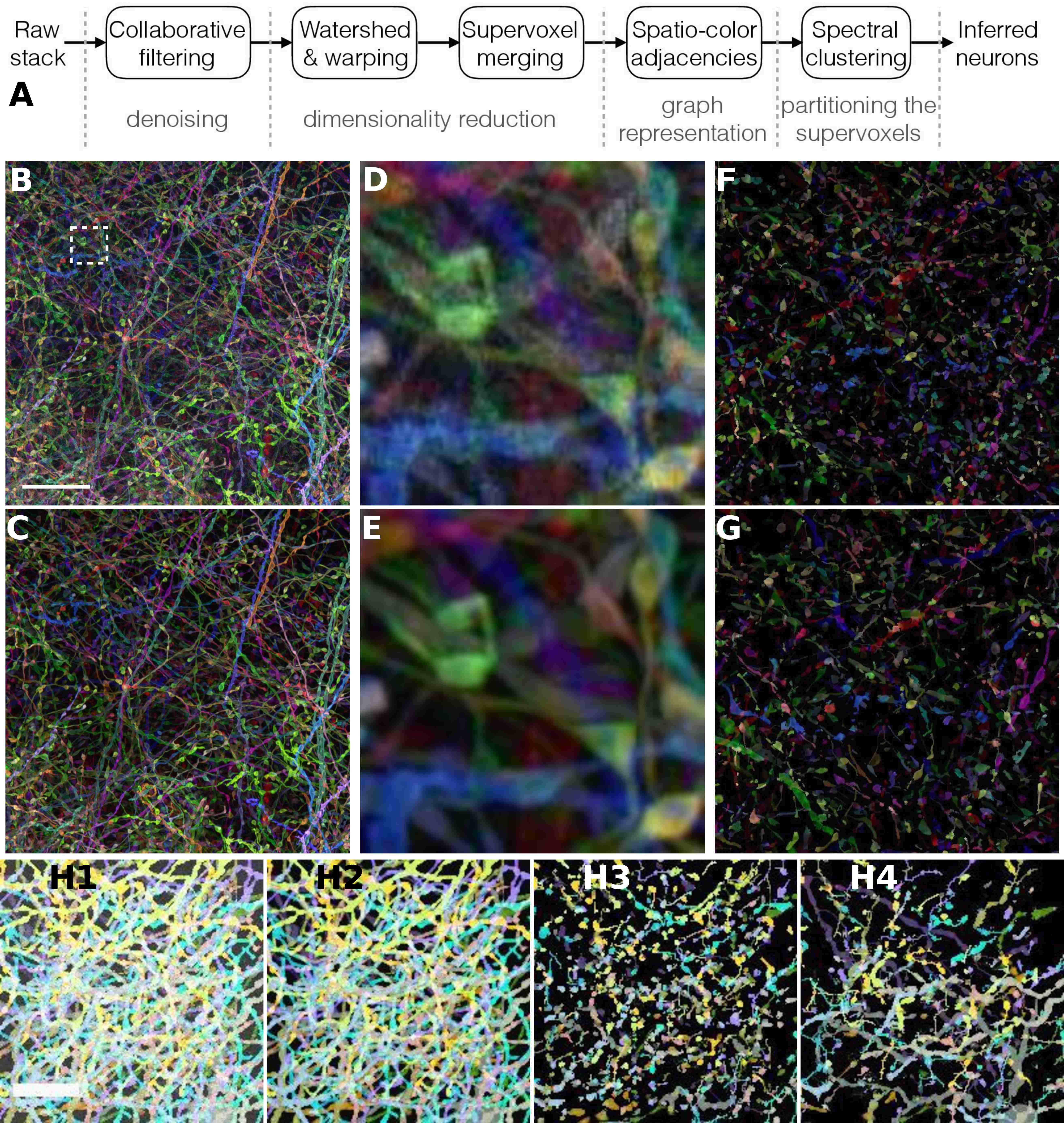}
  \end{minipage}\hfill
  \begin{minipage}[r]{0.18\textwidth}
    \caption{Best viewed digitally. {\bf a,} A schematic of the processing steps {\bf b,} Max. intensity projection of a raw Brainbow image {\bf c,} Max. intensity projection of the denoised image {\bf d,} A zoomed-in version of the patch indicated by the dashed square in {\bf b}. {\bf e,} The corresponding denoised image. {\bf f,} One-third of the supervoxels in the top-left quadrant (randomly chosen). {\bf g,} Same as {\bf f} after the merging step. {\bf h1-h4,} Same as {\bf b,c,f,g} for simulated data. Scale bars, $20 \mu m$.}\label{pipeline2}
  \end{minipage}              
\end{figure}

\vspace{-.3cm}
Voxel colors within a neurite can drift along the neurite, exhibit high frequency variations, and differ between the membrane and the cytoplasm when the expressed fluorescent protein is membrane-binding (Fig.~\ref{noiseSources}). Collaborative filtering generates an extra dimension consisting of similar patches within the stack, and applies filtering in this extra dimension rather than the physical dimensions. We use the BM4D denoiser~\cite{maggioni2013nonlocal} on individual channels of the datasets, assuming that the noise is Gaussian. Figure~\ref{pipeline2} demonstrates that the boundaries are preserved in the denoised image.

\vspace{-.3cm}
\subsection{Dimensionality reduction}
\label{dimensionalityReduction}
\vspace{-.3cm}
We make two basic observations to reduce the size of the dataset: (i) Voxels expressing fluorescent proteins form the foreground, and the dark voxels form the much larger background in typical Brainbow settings. (ii) The basic promise of Brainbow suggests that nearby voxels within a neurite have very similar colors. Hence, after denoising, there must be many topologically connected voxel sets that also have consistent colors.

The watershed transform~\cite{meyer1994topographic} considers its input as a topographic map and identifies regions associated with local minima (``catchment basins'' in a flooding interpretation of the topographic map). It can be considered as a {\em minimum spanning forest} algorithm, and obtained in linear time with respect to the input size~\cite{meyer1994minimum,cousty2009watershed}. For an image volume $V=V(x,y,z,c)$, we propose to calculate the topographical map $T$ (disaffinity map) as
\begin{equation}
\label{topoMap}
T(x,y,z)=\max_{t\in\{x,y,z\}}\max_c |G_t(x,y,z,c)|,
\end{equation}
where $x$, $y$, $z$ denote the spatial coordinates, $c$ denotes the color coordinate, and $G_x$, $G_y$, $G_z$ denote the spatial gradients of $V$ (nearest neighbor differencing). That is, {\em any} edge with significant deviation in {\em any} color channel will correspond to a ``mountain'' in the topographic map. A flooding parameter, $f$, assigns the local minima of $T$ to catchment basins, which partition $V$ together with the boundary voxels. We assign the boundaries to neighboring basins based on color proximity. The background is the largest and darkest basin. We call the remaining objects supervoxels~\cite{ren2003learning, kim2014space}. Let $F$ denote the binary image identifying all of the foreground voxels. 

Objects without interior voxels (e.g., single-voxel thick dendritic segments) may not be detected by Eq.~\ref{topoMap} (Supp. Fig.~1). We recover such ``bridges'' using a topology-preserving warping (in this case, only shrinking is used.) of the thresholded image stack into $F$~\cite{bertrand1994new, jain2010boundary}:
\begin{equation}
B = \mathcal{W}(I_\theta, F),
\end{equation}
where $I_\theta$ is binary and obtained by thresholding the intensity image at $\theta$. $\mathcal{W}$ returns a binary image $B$ such that $B$ has the same topology as $I_\theta$ and agrees with $F$ as much as possible. Each connected component of $B \land \bar{F}$ (foreground of $B$ and background of $F$) is added to a neighboring supervoxel based on color proximity, and discarded if no spatial neighbors exist (Supp. Text).

We ensure the color homogeneity within supervoxels by dividing non-homogeneous supervoxels (e.g., large color variation across voxels) into connected subcomponents based on color until the desired homogeneity is achieved (Supp. Text). We summarize each supervoxel's color by its mean color.

We apply local heuristics and spatio-color constraints iteratively to further reduce the data size and demix overlapping neurons in voxel space (Fig.~\ref{pipeline2}f,g and Supp. Text). Supp. Text provides details on the parallelization and complexity of these steps and the method in general.

\vspace{-.3cm}
\subsection{Clustering the supervoxel set}
\label{clustering}
\vspace{-.3cm}
We consider the supervoxels as the nodes of a graph and express their spatio-color similarities through the existence (and the strength) of the edges connecting them, summarized by a highly sparse adjacency matrix. Removing edges between supervoxels that aren't spatio-color neighbors avoids spurious links. However, this procedure also removes many genuine links due to high color variability (Fig.~\ref{noiseSources}). Moreover, it cannot identify disconnected segments of the same neuron (e.g., due to limited field-of-view). Instead, we adjust the spatio-color neighborhoods based on the ``reliability'' of the colors of the supervoxels. Let $S$ denote the set of supervoxels in the dataset. We define the sets of reliable and unreliable supervoxels as $S_r=\{s\in S: n(s)>t_s, h(s)<t_d\}$ and $S_u=S\setminus S_r$, respectively, where $n(s)$ denotes the number of voxels in $s$, $h(s)$ is a measure of the color heterogeneity (e.g., the maximum difference between intensities across all color channels), $t_s$ and $t_d$ are the corresponding thresholds.

We describe a graph $G=(V,E)$, where $V$ denotes the vertex set (supervoxels) and $E=E_s\cup E_c\cup E_{\bar{s}}$ denotes the edges between them:
\begin{align}
E_s&=\{ (ij): \delta_{ij}<\epsilon_s, i\neq j \} \nonumber\\
E_c&=\{ (ij): s_i,s_j\in S_r, d_{ij}<\epsilon_c,  i\neq j \} \nonumber\\
E_{\bar{s}}&=\{ (ij), (ji): s_i\in S_u, (ij)\notin E_s, O_i(j)<k_{\mathrm{min}}-K_i, i\neq j \},
\end{align}
where $\delta_{ij}$, $d_{ij}$ are the spatial and color distances between $s_i$ and $s_j$, respectively. $\epsilon_s$ and $\epsilon_c$ are the corresponding maximum distances. An unreliable supervoxel with too few spatial neighbors is allowed to have up to $k_{\mathrm{min}}$ edges via proximity in color space. Here, $O_i(j)$ is the order of supervoxel $s_j$ in terms of the color distance from supervoxel $s_i$, and $K_i$ is the number of $\epsilon_s$-spatial neighbors of $s_i$. (Note the symmetric formulation in $E_{\bar{s}}$.) Then, we construct  the adjacency matrix as
\begin{equation}
A(i,j) = \left\{
\begin{array}{ll}
e^{-\alpha d_{ij}^2}, & (ij)\in E \\
0, & \mbox{otherwise}
\end{array}\right.
\end{equation}
where $\alpha$ controls the decay in affinity with respect to distance in color. We use k-d tree structures to efficiently retrieve the color neighborhoods~\cite{bentley1975multidimensional}. Here, the distance between two supervoxels is $\min_{v\in V,u\in U}D(v,u)$, where $V$ and $U$ are the voxel sets of the two supervoxels and $D(v,u)$ is the Euclidean distance between voxels $v$ and $u$.

A classical way of partitioning graph nodes that are nonlinearly separable is by minimizing a function (e.g., the sum or the maximum) of the edge weights that are severed during the partitioning~\cite{wu1993optimal}. Here, we use the normalized cuts algorithm~\cite{ng2002spectral, von2007tutorial} with two simple modifications: the $k$-means step is weighted by the sizes of the supervoxels and initialized by a few iterations of $k$-means clustering of the supervoxel colors only (Supp. Text). The resulting clusters partition the image stack (together with the background), and represent a segmentation of the individual neurons within the image stack. An estimate of the number of neurons can be obtained from a Dirichlet process mixture model~\cite{kurihara2007collapsed}. While this estimate is often rough~\cite{miller2013simple}, the segmentation accuracy appears resilient to imperfect estimates (Fig.~\ref{simQuantification}c).

\vspace{-.3cm}
\subsection{Simulating Brainbow tissues}
\label{simulation}
\vspace{-.3cm}
We create basic simulated Brainbow image stacks from volumetric reconstructions of single neurons (Algorithm~\ref{simulateAlgo}). For simplicity, we model the neuron color shifts by a Brownian noise component on the tree, and the background intensity by a white Gaussian noise component (Supp. Text).

We quantify the segmentation quality of the voxels using the adjusted Rand index (ARI), whose maximum value is $1$ (perfect agreement), and expected value is $0$ for random clusters~\cite{hubert1985comparing}. (Supp. Text)

\begin{algorithm}
\small
\caption{Brainbow image stack simulation}
\label{simulateAlgo}
\begin{algorithmic}[1]
\Require number of color channels $C$, set of neural shapes $S=\{n_i\}_i$, stack (empty, 3d space + color), background noise variability $\sigma_1$, neural color variability $\sigma_2$, $r$, saturation level $M$
\For{$n_i\in S$}
\State Shift and rotate neuron $n_i$ to minimize overlap with existing neurons in the stack
\State Generate a uniformly random color vector $v_i$ of length $C$
\State Identify the connected components of $c_{ij}$ of $n_i$ within the stack
\For{$c_{ij} \in \{c_{ij}\}_j$}
\State Pre-assign $v_i$ to $r\%$ of the voxels of $c_{ij}$ 
\State $C$-dimensional random walk on $c_{ij}$ with steps $\mathcal{N}(0,\sigma_1^2 \bf{I})$ (Supp. Text)
\EndFor
\State Add neuron $n_i$ to the stack (with additive colors for shared voxels)
\EndFor
\State{Add white noise to each voxel generated by $\mathcal{N}(0,\sigma_2^2 \bf{I})$} 
\If{brightness exceeds $M$}
\State{Saturate at $M$}
\EndIf
\State \Return stack
\end{algorithmic}
\end{algorithm}

\begin{figure}
  \begin{minipage}[l]{0.72\textwidth}
    \includegraphics*[viewport = 0 0 800 650, width=\textwidth]{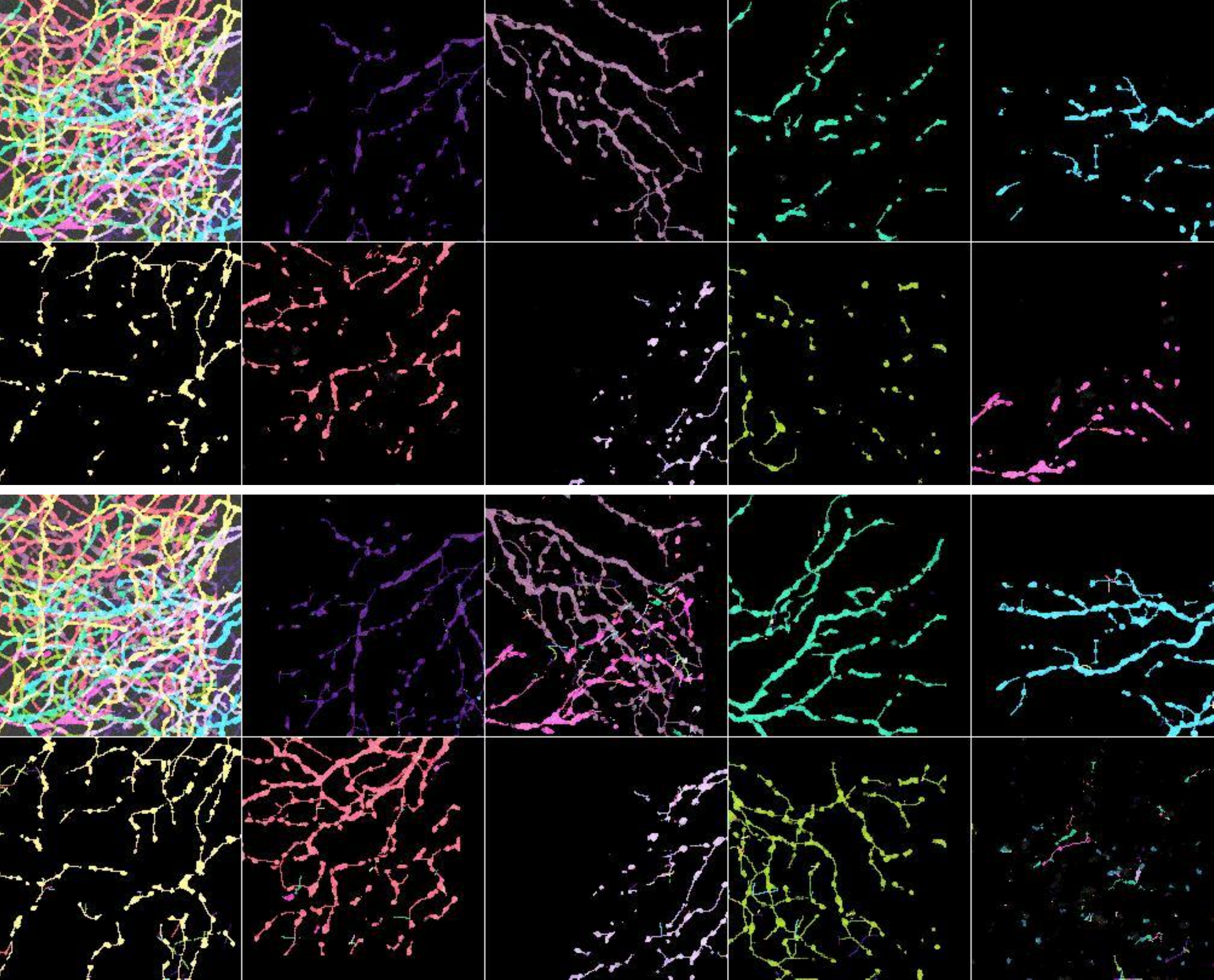}
  \end{minipage}\hfill
  \begin{minipage}[r]{0.27\textwidth}
    \caption{{\bf Segmentation of a simulated Brainbow image stack. Adjusted Rand index of the foreground is $0.80$. Pseudo-color representation of 4-channel data.} {\bf Top:} maximum intensity projection of the ground truth. Only the supervoxels that are occupied by a single neuron are shown. {\bf Bottom:} maximum intensity projection of the reconstruction. The top-left corners show the whole image stack. All other panels show the maximum intensity projections of the supervoxels assigned to a single cluster (inferred neuron).} \label{simVisualization}
  \end{minipage}
\end{figure}

\begin{figure}
\centering
\includegraphics*[viewport = 0 0 2850 750, width=\textwidth]{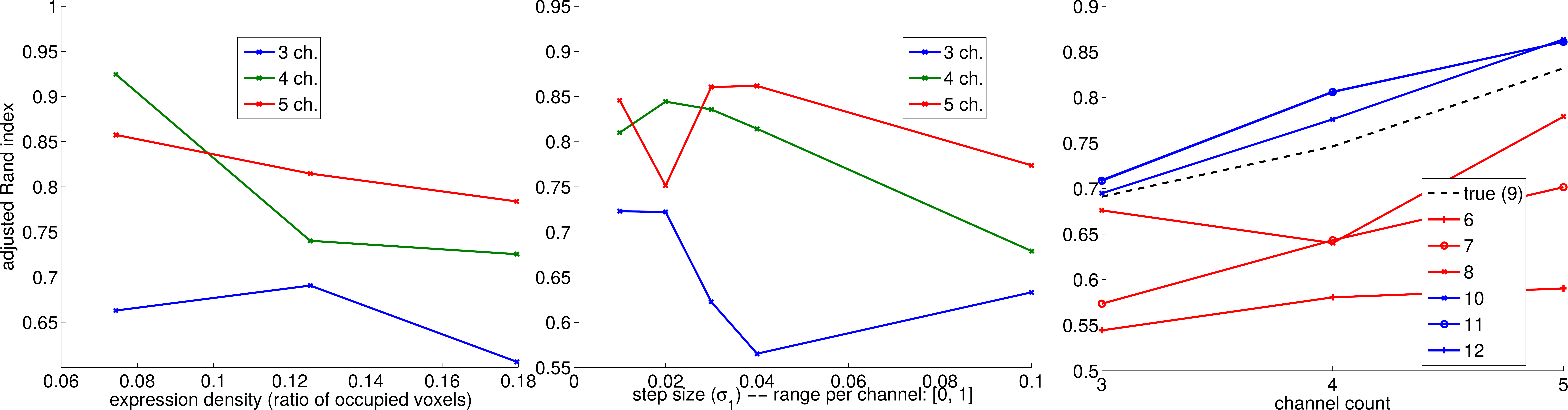}
\caption{{\bf Segmentation accuracy of simulated data} {\bf a,} Expression density (ratio of voxels occupied by at least one neuron) vs. ARI. {\bf b,} $\sigma_1$ (Algorithm~\ref{simulateAlgo}) vs. ARI. {\bf c,} Channel count vs. ARI for a $9$-neuron simulation, where $K\in [6,12]$. ARI is calculated for the foreground voxels. See Supp. Fig.~7 for ARI values for all voxels.}
\label{simQuantification}
\end{figure}

\begin{figure}
\centering
\includegraphics*[viewport = 0 0 5125 7175, width=\textwidth]{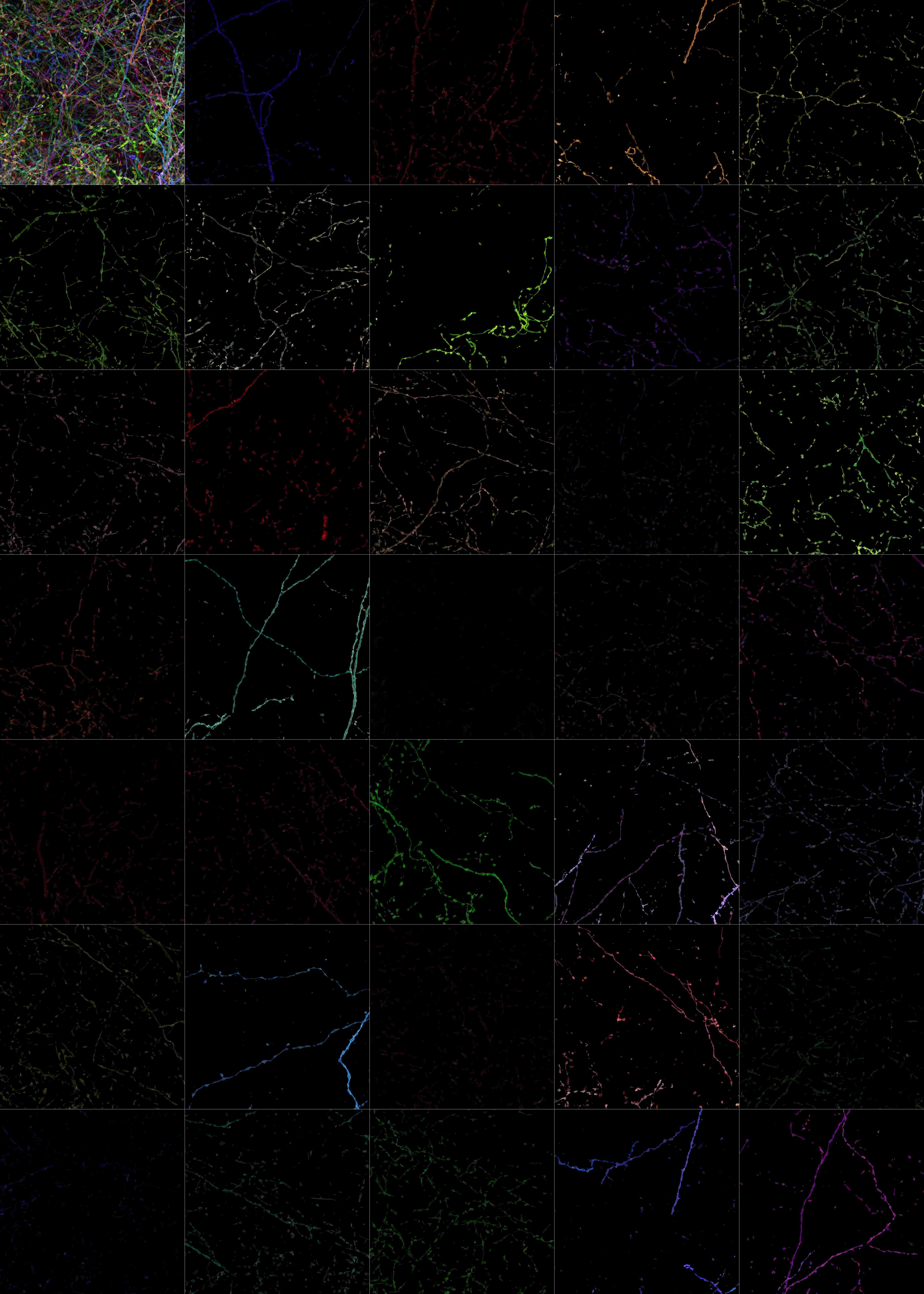}
\caption{{\bf Segmentation of a Brainbow stack -- best viewed digitally. Pseudo-color representation of 4-channel data.} The physical size of the stack is $102\mu\times 102\mu\times 68\mu$. The top-left corner shows the maximum intensity projection of the whole image stack, all other panels show the maximum intensity projections of the supervoxels assigned to a single cluster (inferred neuron).}
\label{dcai_bb_4channel}
\end{figure}

\vspace{-.3cm}
\section{Datasets}
\vspace{-.3cm}
To simulate Brainbow image stacks, we used volumetric single neuron reconstructions of mouse retinal ganglion cells in Algorithm~\ref{simulateAlgo}. The dataset is obtained from previously published studies~\cite{sumbul2014genetic,sumbul2014automated}. Briefly, the voxel size of the images is $0.4\mu\times 0.4\mu\times 0.5\mu$, and the field of view of individual stacks is $320\mu\times 320\mu\times 70\mu$ or larger. We evaluate the effects of different conditions on a central portion of the simulated image stack.

Both real datasets are images of the mouse hippocampal tissue. The first dataset has $1020\times 1020\times 225$ voxels (voxel size: $0.1\times 0.1\times 0.3\mu^3$), and the tissue was imaged at $4$ different frequencies (channels). The second dataset has $1080\times 1280\times 134$ voxels with an effective voxel size of $70\times 70\times 40nm$, where the tissue was $4\times$ linearly expanded~\cite{chen2015expansion}, and imaged at $3$ different channels. The Brainbow constructs were delivered virally, and approximately $5\%$ of the neurons express a fluorescence gene.

\vspace{-.3cm}
\section{Results}
\vspace{-.3cm}
Parameters used in the experiments are reported in Supp. Text.

Fig.~\ref{noiseSources}b, d, and~e depict the variability of color within individual neurites in a single slice and through the imaging plane. Together, they demonstrate that the voxel colors of even a small segment of a neuron's arbor can occupy a significant portion of the dynamic range in color with the state-of-the-art Brainbow data. Fig.~\ref{noiseSources}c-e show that collaborative denoising removes much of this noise while preserving the edges, which is crucial for segmentation. Fig.~\ref{pipeline2}b-e and~h demonstrate a similar effect on a larger scale with real and simulated Brainbow images.

Fig.~\ref{pipeline2} shows the raw and denoised versions of the $1020\times 1020\times 225$ image, and a randomly chosen subset of its supervoxels (one-third). The original set had $6.2\times 10^4$ supervoxels, and the merging routine decreased this number to $3.9\times 10^4$. The individual supervoxels grew in size while avoiding mergers with supervoxels of different neurons. This set of supervoxels, together with a (sparse) spatial connectivity matrix, characterizes the image stack. Similar reductions are obtained for all the real and simulated datasets.

Fig.~\ref{simVisualization} shows the segmentation of a simulated $200\times 200\times 100$ (physical size: $80\mu\times 80\mu\times 50\mu$) image patch. (Supp. Fig.~2 shows all three projections, and Supp. Fig.~3 shows the density plot through the $z$-axis.) In this particular example, the number of neurons within the image is $9$, $\sigma_1=0.04$, $\sigma_2=0.1$, and the simulated tissue is imaged using $4$ independent channels. Supp. Fig.~4 shows a patch from a single slice to visualize the amount of noise. The segmentation has an adjusted Rand index of $0.80$ when calculated for the detected foreground voxels, and $0.73$ when calculated for all voxels. (In some cases, the value based on all voxels is higher.) The ground truth image displays only those supervoxels all of whose voxels belong to a single neuron. The bottom part of Fig.~\ref{simVisualization} shows that many of these supervoxels are correctly clustered to preserve the connectivity of neuronal arbors. There are two important mistakes in clusters 4 (merger) and 9 (spurious cluster). These are caused by aggressive merging of supervoxels (Supp. Fig.~5), and the segmentation quality improves with the inclusion of an extra imaging channel and more conservative merging (Supp. Fig.~6). We plot the performance of our method under different conditions in Fig.~\ref{simQuantification} (and Supp. Fig.~7). We set the noise standard deviation to $\sigma_1$ in the denoiser, and ignored the contribution of $\sigma_2$. Increasing the number of observation channels improves the segmentation performance. The clustering accuracy degrades gradually with increasing neuron-color noise ($\sigma_1$) in the reported range (Fig.~\ref{simQuantification}b). The accuracy does not seem to degrade when the cluster count is mildly overestimated, while it decays quickly when the count is underestimated (Fig.~\ref{simQuantification}c).

Fig.~\ref{dcai_bb_4channel} displays the segmentation of the $1020\times 1020\times 225$ image. While some mistakes can be spotted by eye, most of the neurites can be identified and simple tracing tools can be used to obtain final skeletons/segmentations~\cite{longair2011simple, peng2010v3d}. In particular, the identified clusters exhibit homogeneous colors and dendritic pieces that either form connected components or miss small pieces that do not preclude the use of those tracing tools. Some clusters appear empty while a few others seem to comprise segments from more than one neuron, in line with the simulation image (Fig.~\ref{simulation}).

Supp. Fig.~8 displays the segmentation of the $4\times$ expanded, $1080\times 1280\times 134$ image. While the two real datasets have different characteristics and voxel sizes, we used essentially the same parameters for both of them throughout denoising, supervoxelization, merging, and clustering (Supp. Text). Similar to Fig.~\ref{dcai_bb_4channel}, many of the processes can be identified easily. On the other hand, Supp. Fig.~8 appears more fragmented, which can be explained by the smaller number of color channels (Fig.~\ref{simQuantification}).

\vspace{-.3cm}
\section{Discussion}
\vspace{-.3cm}
Tagging individual cells with (pseudo)colors stochastically is an important tool in biological sciences. The versatility of genetic tools for tagging synapses or cell types and the large field-of-view of light microscopy positions multispectral labeling as a complementary approach to electron microscopy based, small-scale, dense reconstructions~\cite{takemura2013visual}. However, its use in neuroscience has been limited due to various sources of variability in expression. Here, we demonstrate that automated segmentation of neurons in such image stacks is possible. Our approach considers both accuracy and scalability as design goals.

The basic simulation proposed here (Algo.~\ref{simulateAlgo}) captures the key aspects of the problem and may guide the relevant genetics research. Yet, more detailed biophysical simulations represent a valuable direction for future work. Our simulations suggest that the segmentation accuracy increases significantly with the inclusion of additional color channels, which coincides with ongoing experimental efforts~\cite{zador2012sequencing,lee2014highly,chen2015spatially}. We also note that color constancy of individual neurons plays an important role both in the accuracy of the segmentation (Fig.~\ref{simQuantification}) and the supervoxelized problem size.

While we did not focus on post-processing in this paper, basic algorithms (e.g., reassignment of small, isolated supervoxels) may improve both the visualization and the segmentation quality. Similarly, more elaborate formulations of the adjacency relationship between supervoxels can increase the accuracy. Finally, supervised learning of this relationship (when labeled data is present) is a promising direction, and our methods can significantly accelerate the generation of training sets.

\section{Acknowledgments}
The authors thank Suraj Keshri and Min-hwan Oh (Columbia University) for useful conversations.

Funding for this research was provided by ARO MURI W911NF-12-1-0594, DARPA N66001- 15-C-4032 (SIMPLEX), and a Google Faculty Research award; in addition, this work was supported by the Intelligence Advanced Research Projects Activity (IARPA) via Department of Interior/ Interior Business Center (DoI/IBC) contract number D16PC00008. The U.S. Government is authorized to reproduce and distribute reprints for Governmental purposes notwithstanding any copyright annotation thereon. Disclaimer: The views and conclusions contained herein are those of the authors and should not be interpreted as necessarily representing the official policies or endorsements, either expressed or implied, of IARPA, DoI/IBC, or the U.S. Government.

\newpage
\setlength{\bibsep}{.4ex}
\begin{small}
\bibliographystyle{unsrtnat}
\bibliography{bb}
\end{small}
\end{document}


\maketitle
\section*{Supplementary Text}
\subsection*{Parameter choices and feature representations}

Similar results are obtained around a neighborhood of these suggested values. Unless otherwise noted, the same parameter values were used for all three reported experiments: two real hippocampal datasets acquired by different labs and under different conditions, and a set of simulated retinal datasets with different parameters. The subsection of the main text that refers to these parameters are indicated in square brackets.

{\bf Color features:} For every color triplet, we obtain the L-u-v representation~\citep{schanda2007colorimetry}, and calculate the top $C$ principal components of the concatenated L-u-v representations, where $C$ is the number of color channels. If the data has more than $3$ channels, for affinity calculations between neighboring supervoxels, we normalize the colors before the L-u-v transformation. [Dimensionality reduction]

{\bf Supervoxel reliability:} $t_s = 50$, $t_d = 0.5$ (before L-u-v transformation) [Clustering the supervoxel set]
 
{\bf Edge set parameters:} $\epsilon_s = \sqrt{3}$ ($26$-neighborhood for isotropic data), $\epsilon_c = 20 \times \sqrt{C/4}$ (by inspecting typical color radius within individual neurons and manual adjustment), $k_{\rm min}=5$. [Clustering the supervoxel set]

{\bf Edge strength decay:} $\alpha=2\times 10^{-3}$ (by inspecting typical color radius and manual adjustment) [Clustering the supervoxel set]

{\bf Flooding parameter for watershed:} $f=0.01$ with 26-neighborhood. (This affects computation time more than quality because subdividing via the maximum color perimeter can catch inhomogeneous supervoxels.) [Dimensionality reduction]

{\bf Maximum color perimeter for supervoxel homogeneity:} (Supp. Algo.~1) $p=0.5$ for each channel when the intensity is in $[0,1]$ (by inspecting data -- see Fig.~1). [Dimensionality reduction]

{\bf Image thresholding for warping:} $\theta=0.1 \times \sqrt{C/4}$ before L-u-v transformation (for the expansion microscopy data, $\theta=0.2$) [Dimensionality reduction]

{\bf Noise standard deviation for denoising:} $\sigma=1/8$ when the intensity is in $[0,1]$ for individual channels. [Denoising the image stack]

{\bf Cluster (neuron) counts:} For the dataset in Fig.~5, the mixture model~\citep{kurihara2007collapsed} suggested $52$ clusters based on the colors of the supervoxels. The same routine returned $29$ clusters when run on $\frac{1}{5}$ of the supervoxels. We chose $K=34$ for a compact presentation. For the dataset in Supp. Fig.~8, we used $K=19$, which is what the mixture model suggested based on the colors of the supervoxels. [Clustering the supervoxel set]

\subsection*{Spatial distance calculation}
The spatial distance between two supervoxels is calculated as $\min_{v\in V_1,u\in V_2}D(v,u)$, where $V_1$ and $V_2$ are the voxel sets of the two supervoxels and $D(v,u)$ is the Euclidean distance between voxels $v$ and $u$. Only the boundary voxels need to be considered, and extremal values in each coordinate are used to identify many supervoxel pairs farther than $\epsilon_s$ without exact calculation over the voxels. Only the spatial distances between nearby supervoxels need to be computed.

\subsection*{Color-based subdivision of supervoxels}
Let the $n\times C$ matrix $V_i$ denote the colors of all $n$ voxels of the supervoxel $s_i$. Supp. Algo.~1 divides the supervoxels into smaller supervoxels until the desired homogeneity is achieved.
\begin{algorithm}
\small
\caption{Subdivide supervoxels}
\begin{algorithmic}[1]
\label{subdivideSupervoxels}
\Require $S=\{s_i\}_i$ (set of supervoxels), $\{V_i\}_i$, $p_{\rm max}$ (threshold)
\State $S_{\rm new}=\{\}$
\For{$s_i\in S$}
\State $p = \max_{c \in C} \max(V_i(:,c))-\min(V_i(:,c))$
\If{$p<p_{\rm max}$}
\State Add $s_i$ to $S_{\rm new}$
\Else
\State Divide the voxels into $2$ sets $T_1$ and $T_2$ based on color (e.g., using $k$-means, hierarchical clustering, etc.)
\State Add the connected components of $T_1$ and $T_2$ to $S$
\EndIf
\State Remove $s_i$ from $S$
\EndFor
\State \Return $S_{\rm new}$
\end{algorithmic}
\end{algorithm}

\subsection*{Simulation data}
RGC arbors stratify in the retina, distributing their dendritic length within a slab. To achieve denser simulations, we did not shift the neurons much in $z$ (The density numbers calculated in Fig.~4 are obtained by considering the $[35\mu m, 50\mu m]$ region in Supp. Fig.~3.) We obtain simulated stacks by varying the expression density ($|S|\in\{5,9,13\}$), the channel count ($C\in\{3,4,5\}$), the neuron color consistency ($\sigma_1\in\{0.01,0.02,0.03,0.04, 0.1\}$), and the background noise ($\sigma_2\in\{0.05,0.1\}$).

The random walk on a connected component assigns a color to a voxel by (i) calculating the mean color of the neighboring voxels that were previously visited, and (ii) adding the random noise step to this mean value.

\subsection*{Adjusted Rand index}
We quantify the segmentation quality of the voxels of the simulated dataset using the adjusted Rand index. The Rand index is a measure of the element pairs on which two partitions $P$ and $\hat{P}$ of the same set with $N$ elements agree: $R(P,\hat{P})=1-\binom{N}{2}\sum_{i<j}|\delta(l_i,l_j)-\delta(\hat{l_i},\hat{l_j})|$, where $l_i$ ($\hat{l_i}$) denotes the label of element $i$ according to $P$ ($\hat{P}$), and $\delta$ is the indicator function. ($\delta(l_i,l_j)=1$ if $l_i=l_j$, $\delta(l_i,l_j)=0$ otherwise.) The adjusted Rand index corrects for chance, has a more sensitive dynamic range, and is defined as $A=(R-E)/(M-E)$, where $E$ is the expected value of the index and $M$ is the maximum value of the index, based on the number of elements in individual segments. Its maximum value is $1$ (perfect agreement), and the expected value of the index is $0$ for random clusters.

For the foreground based calculation, only the voxels that are assigned to the foreground after the watershed transform and warping are considered. For the image based calculation, all voxels are considered and the background is treated as a separate object.

\subsection*{Merging supervoxels}
We apply local heuristics and spatio-color constraints iteratively to further reduce the data size and demix overlapping neurons in voxel space (Fig.~2): (i) supervoxels occupied by more than one neuron are detected and demixed by monitoring the improvement in non-negative least squares fit quality. (Supp. Algo~2.) (ii) neighboring supervoxels with similar colors and orientations, supervoxels with single spatial neighbors, and supervoxels all of whose neighbors have similar colors are merged. (iii) supervoxels that are spatial neighbors and that are assigned to the same cluster by an overclustering color $k$-means routine are merged. We implement (iii) to run in parallel over subgraphs of the full graph for scalability. A rough estimate of the number of neurons required by the oversegmentation routine is obtained by a Dirichlet process mixture model~\citep{kurihara2007collapsed}. The $k$-means algorithm uses a multiple of this rough estimate. Note that only a rough estimate~\citep{miller2013simple} is needed because of oversegmentation (Supp Algo.~3). This algorithm can be implemented to run in parallel over subgraphs of the full dataset.
\begin{algorithm}
\small
\caption{Demixing of supervoxels}
\begin{algorithmic}[1]
\label{demixSupervoxels}
\Require $S=\{s_i\}_i$, $V$ (matrix of normalized supervoxel colors), $A$ (spatial affinity matrix), $M$ (maximum size), $\Delta$ (maximum color distance), $f$ (improvement factor)
\For{$s_i\in S$}
\If{$s_i$ has less than $M$ voxels}
\State Retain the neighbors that have neighbors with color distance less than $\Delta$
\If{$s_i$ has more than one spatial neighbors}
\If{the minimum color distance between $s_i$ and its neighbors is larger than $\Delta$}
\State Initialize $r=||V(i,:)||_2^2$, $P=(0, 0)$
\For{each neighbor pair $(i_1, i_2)$}
\State $t=\min_x ||V([i_1, i_2], :)x-V(i,:)||_2^2$ subject to $x\geq 0$
\If{$t<r$}
\State $r=t$, $P=(i_1, i_2)$
\EndIf
\EndFor
\If{$r<(\Delta/f)^2$}
\State Assign the voxels of $s_i$ to both of $s_{P(1)}$ and $s_{P(2)}$
\State Update the spatial affinities of $s_{P(1)}$ and $s_{P(2)}$ in $A$ accordingly
\State Remove $s_i$ from $S$, $V$, and $A$
\EndIf
\EndIf
\EndIf
\EndIf
\EndFor
\State \Return $S$, $V$, $A$
\end{algorithmic}
\end{algorithm}

\begin{algorithm}
\small
\caption{Spatio-color merging of supervoxels}
\begin{algorithmic}[1]
\label{mergeSupervoxels}
\Require $S=\{s_i\}_i$, (set of supervoxels) $K$ (rough estimate of the number of clusters), $k$ (oversegmentation factor)
\State $S_{\rm new}=\{\}$
\State Divide $S$ into $kK$ clusters based on the colors of the supervoxels, using $k$-means
\For{$\kappa_1\in\{1,\dots,kK\}$}
\State Find the connected components within the cluster $\kappa_1$
\State Merge the supervoxels within the connected components of that cluster, and add to $S_{\rm new}$
\EndFor
\State \Return $S_{\rm new}$
\end{algorithmic}
\end{algorithm}

\newpage
\begin{figure}
\centering
\includegraphics*[viewport = 0 0 1770 1770, width=\textwidth]{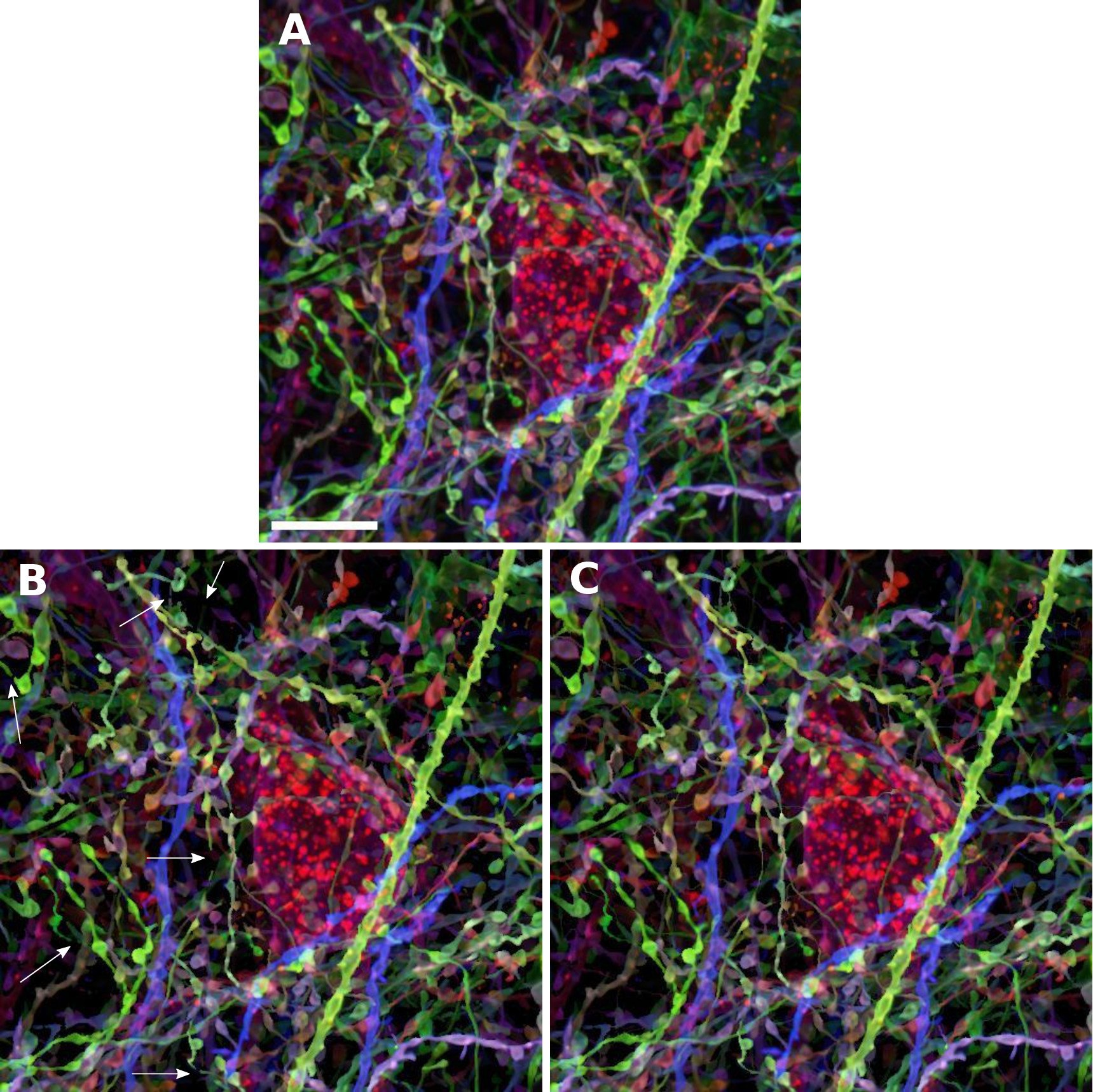}
\caption{{\bf Top:} Maximum intensity projection of a raw Brainbow image. {\bf Bottom left:} Foreground after watershed transform. Arrows point to six different thin dendritic pieces (``bridges'') that were missed. {\bf Bottom right:} Foreground after warping correction. Scale bar, $30\mu m$}
\end{figure}
\begin{figure}
\centering
\includegraphics*[viewport = 0 0 350 350, width=0.9\textwidth]{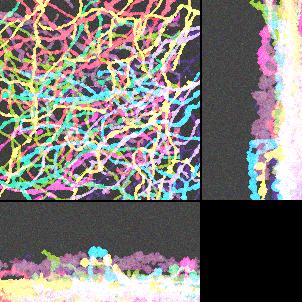}
\caption{{\bf The $z$ (top left), $x$ (top right), and $y$ (bottom left) maximum intensity projections of the raw simulation image shown in Fig. 3. Adjusted Rand index of the segmentation is $0.80$.}}
\end{figure}
\begin{figure}
\centering
\includegraphics*[viewport = 0 150 600 600, width=\textwidth]{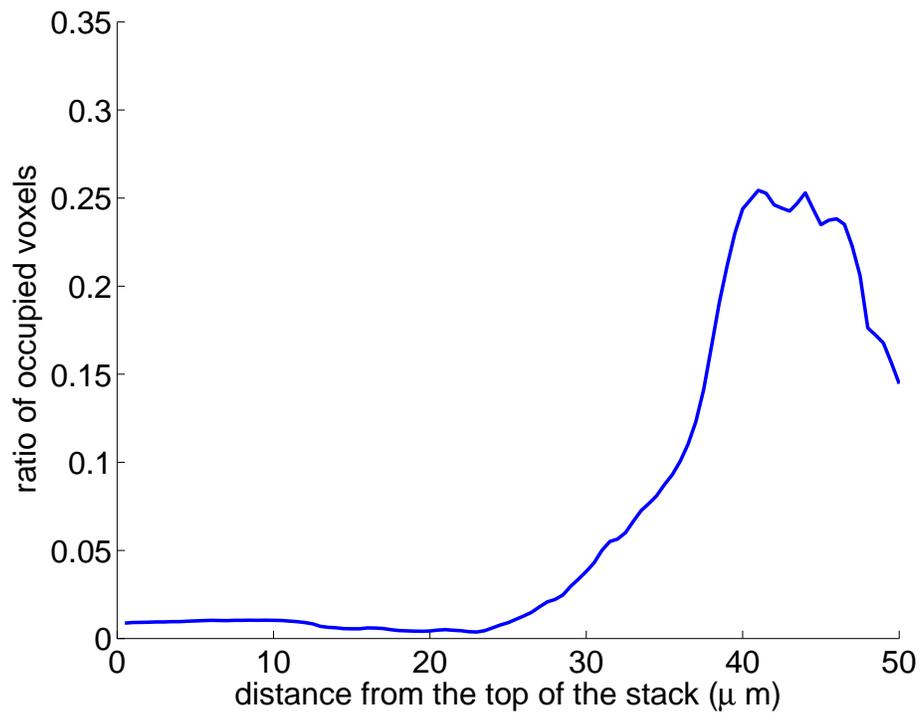}
\caption{{\bf The $z$-profile of the ground truth simulation image with $13$ neurons, showing that a region is preferentially occupied. The range $[0\mu m, 50\mu m]$ corresponds to slices $1$ to $100$ so that most of the neuronal arbors are between slices $70$ and $100$.}}
\end{figure}
\begin{figure}
\centering
\includegraphics*[viewport = 0 0 60 102, width=0.90\textwidth]{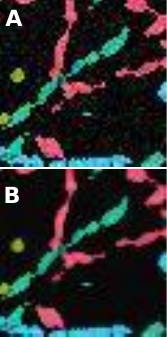}
\caption{{\bf A $60\times 60$ patch from a single slice (slice 90) of the simulation image shown in Fig. 3.} {\bf Top:} raw. {\bf Bottom:} denoised. Physical size: $15\mu m\times 15\mu m$}
\end{figure}

\begin{figure}
\centering
\includegraphics*[viewport = 0 0 160 330, width=0.7\textwidth]{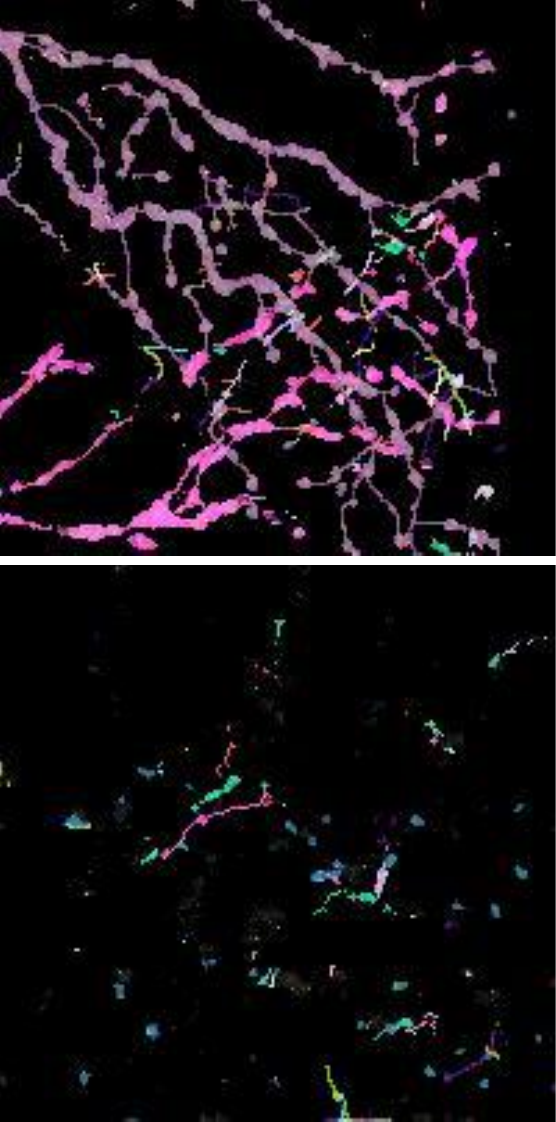}
\caption{{\bf Aggressive merging generates supervoxels with inconsistent colors.} {\bf Top:} Cluster~4 in the bottom part of Fig.~3 of the main text. {\bf Bottom:} Cluster~9 in the bottom part of Fig.~3 of the main text. Close inspection reveals that some of the multi-colored regions comprise single supervoxels.}
\end{figure}

\begin{figure}
\centering
\includegraphics*[viewport = 0 0 1010 402, width=\textwidth]{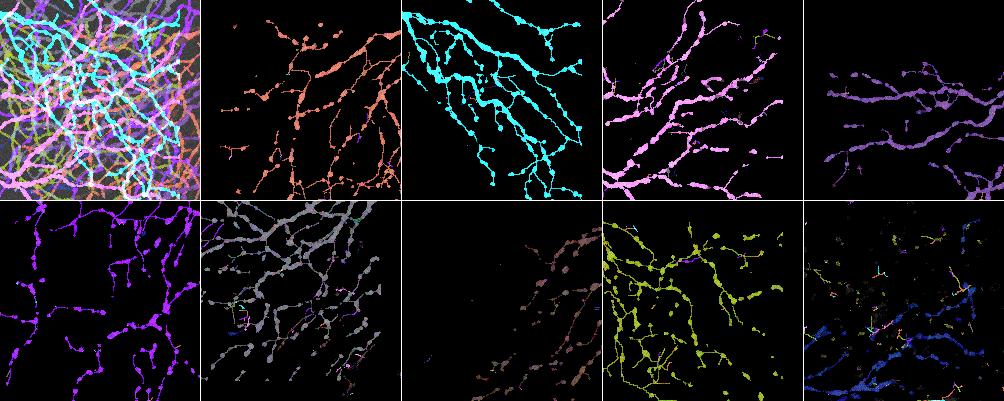}
\caption{ {\bf Segmentation of a simulated Brainbow image stack. Adjusted Rand index of the foreground is $0.87$. Pseudo-color representation of 5-channel data with more conservative supervoxel merging compared to Fig.~3 of the main text.} Maximum intensity projection of the segmentation. The top-left corner shows the whole image stack. All other panels show the maximum intensity projections of the supervoxels assigned to a single cluster (inferred neuron). }
\end{figure}

\begin{figure}
\centering
\includegraphics*[viewport = 0 0 2850 750, width=\textwidth]{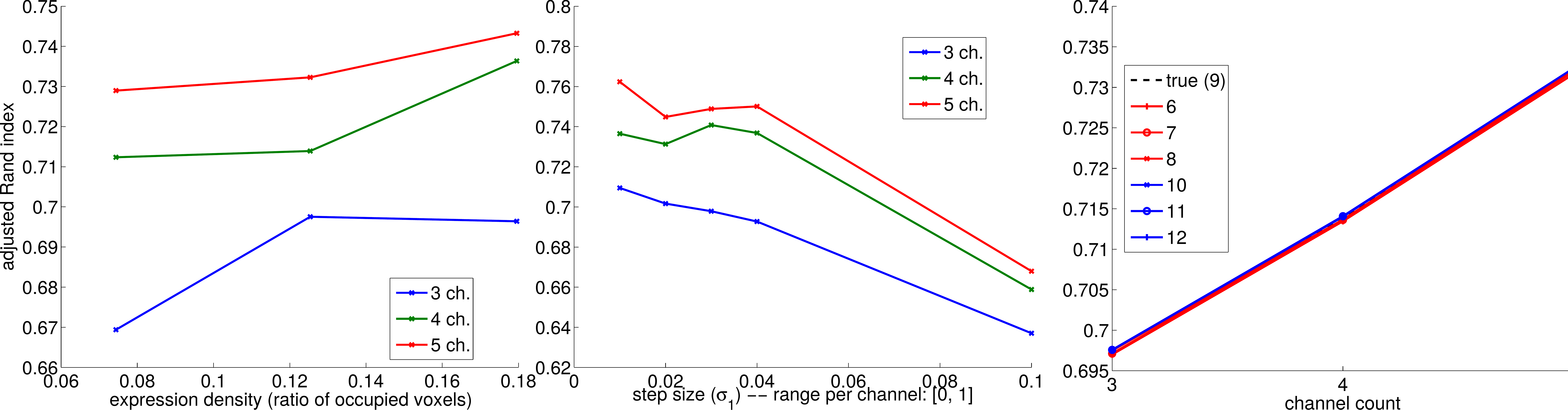}
\caption{{\bf Segmentation accuracy of simulated data} {\bf a,} Expression density (ratio of voxels occupied by at least one neuron) vs. ARI. {\bf b,} $\sigma_1$ vs. ARI. {\bf c,} Channel count vs. ARI for a $9$-neuron simulation, where $K\in [6,12]$. ARI is calculated for all voxels.}
\end{figure}
\clearpage
\begin{figure}
\centering
\includegraphics*[viewport = 0 0 5125 5405, width=\textwidth]{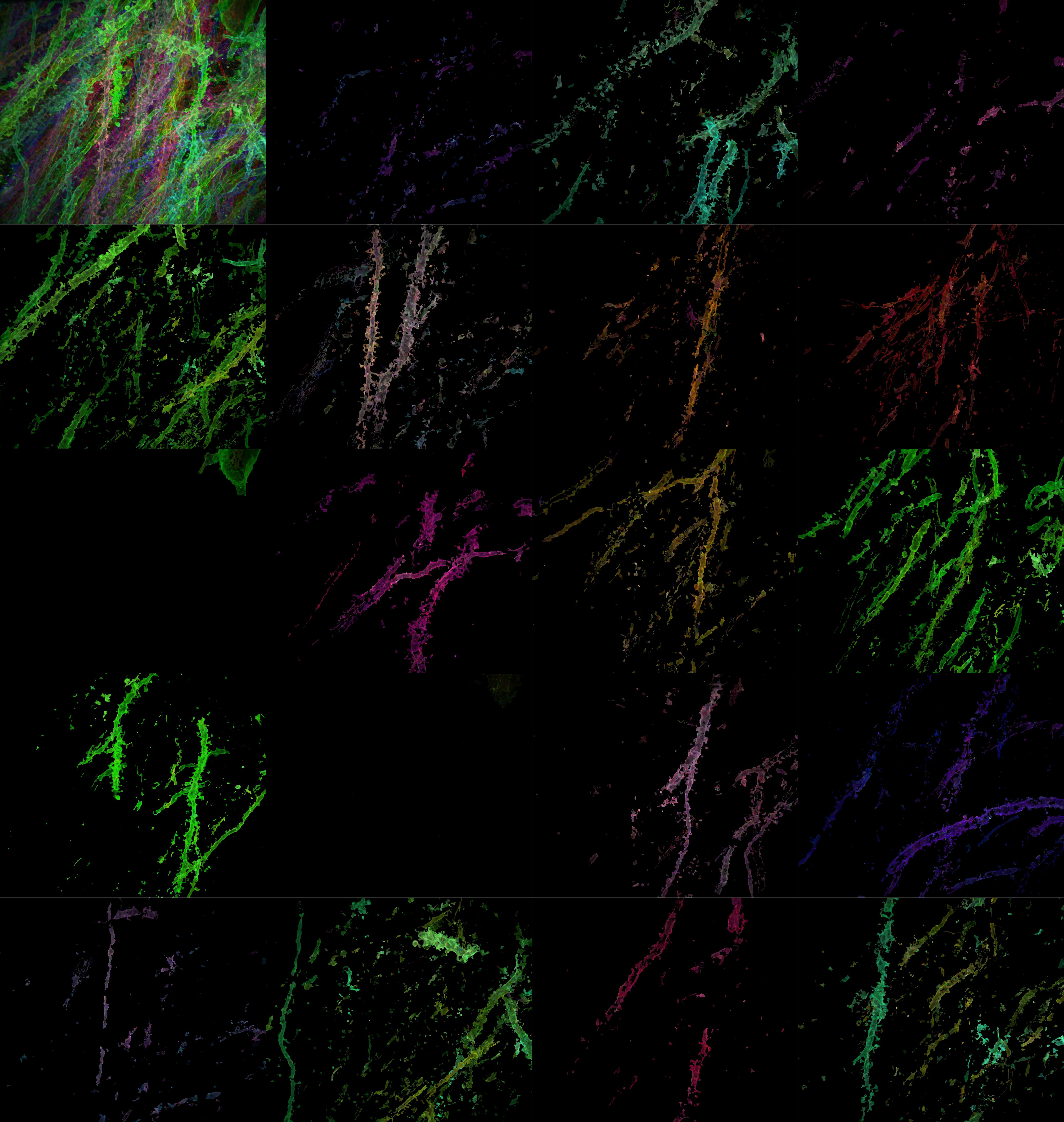}
\caption{{\bf Segmentation of a $4\times$ linearly expanded Brainbow stack -- best viewed digitally.} The physical size of the stack is $90\mu\times 76\mu\times 5\mu$. The top-left corner shows the maximum intensity projection of the whole image stack, all other panels show the maximum intensity projections of the supervoxels assigned to a single cluster (inferred neuron).}
\end{figure}

\subsection*{Parallelization and complexity}
The denoising step (collaborative filtering) parallelizes over substacks (voxels) because the extra dimension is formed by local patches.

The watershed algorithm can also be parallelized over substacks. Moreover, it has a linear run-time with respect to the input size (Main Text).

Similarly, warping (shrinking) can be applied on those substacks. Querying the boundary voxels for flipping at each stage, and ordering them by brightness result in a fast implementation. It has a linear run-time  with respect to the input size. (Simple voxel query can be performed over $3\times 3\times 3$ patches -- See references in main text.)

Supervoxel merging can be performed in parallel on individual substacks. Merging subroutines use local rules to make local merges except for the spatio-color merging step, which uses the $k$-means algorithm.

Similarity calculations require extracting spatial and color neighborhoods of the supervoxels. Spatial distance calculation is discussed above. These value are precalculated. As mentioned in the main text, k-d tree structures are used to retrieve the color neighborhoods efficiently.

Finally, clustering is performed by the normalized cuts algorithm. We use an implicitly restarted block Lanczos method for computing the first few eigenvectors~\citep{baglama2003algorithm}.

\setlength{\bibsep}{.5ex}
\begin{small}
\bibliographystyle{apalike}
\bibliography{bbsupp}
\end{small}